\documentclass[12pt,aps,article]{revtex4}

\usepackage{amsmath}
\usepackage{mathrsfs}
\usepackage{bm}
\usepackage[pdftex]{hyperref}
\usepackage{amsfonts}

\begin{document}
\begin{titlepage}	
\title{Parabolic cylinder functions implemented in Matlab}
\author{E. Cojocaru}
\affiliation{Department of Theoretical Physics, 
Horia Hulubei National Institute of Physics and Nuclear Engineering, Magurele-Bucharest P.O.Box MG-6, 077125 Romania}

\email{ecojocaru@theory.nipne.ro}

\begin{abstract}
Routines for computation of Weber's parabolic cylinder functions and their derivatives are implemented in Matlab for both moderate and great values of the argument. Standard, real solutions are considered. Tables of values are included.
\end{abstract}

\maketitle

\end{titlepage}
\section{Introduction}
The parabolic cylinder functions were introduced by Weber \cite{1} in 1869. Standard solutions to Weber's equation were given by Miller \cite{2} in 1952. These relations are also provided by Abramowitz and Stegun \cite{3}.
There are two standard forms of the Weber's equation,
\begin{equation}
\label{eq:1}
\frac{\mathrm{d}^2 y}{\mathrm{d}x^2}-(\tfrac{1}{4}x^2+a)y=0,
\end{equation}
\begin{equation}
\label{eq:2}
\frac{\mathrm{d}^2 y}{\mathrm{d}x^2}+(\tfrac{1}{4}x^2-a)y=0.
\end{equation}
Equation~(\ref{eq:2}) is obtained from~(\ref{eq:1}) with changes $a$ by $-ia$ and $x$ by $xe^{i\pi/4}$. Thus, if $y(a,x)$ is a solution of~(\ref{eq:1}), then ~(\ref{eq:2}) has solutions: $y(-ia,xe^{i\pi/4}), y(-ia,-xe^{i\pi/4}), y(ia,-xe^{-i\pi/4})$, and $y(ia,xe^{-i\pi/4})$. In the following we consider only real solutions of real equations.

\section{Solutions of equation~(\ref{eq:1})}

\subsection{Standard solutions}

There are two standard solutions of Eq.~(\ref{eq:1}), $U(a,x)$ and $V(a,x)$, both of them expressed in terms of Whittaker's function $D_{-a-\frac{1}{2}}$,
\begin{equation}
\label{eq:3}
U(a,x)=D_{-a-\frac{1}{2}},
\end{equation}
\begin{equation}
\label{eq:4}
V(a,x)=\tfrac{1}{\pi}\Gamma(\tfrac{1}{2}+a)[\sin(\pi a)U(a,x)+U(a,-x)].
\end{equation}
In a more symmetrical notation, these solutions are
\begin{equation}
\label{eq:5}
U(a,x)=D_{-a-\frac{1}{2}}=Y_1 \cos\beta -Y_2 \sin\beta,
\end{equation}
\begin{equation}
\label{eq:6}
V(a,x)=\frac{1}{\Gamma(\tfrac{1}{2}-a)}(Y_1 \sin\beta + Y_2 \cos\beta),
\end{equation}
where
\begin{equation}
\label{eq:7} 
\beta=\pi(\tfrac{a}{2}+\tfrac{1}{4}),
\end{equation}
\begin{equation}
\label{eq:8}
Y_1 =\frac{y_1 \Gamma(\tfrac{1}{4}-\tfrac{a}{2})}{\sqrt\pi 2^{\frac{a}{2}+
\frac{1}{4}}}, \qquad
Y_2=\frac{y_2\Gamma(\tfrac{3}{4}-\tfrac{a}{2})}{\sqrt\pi 2^{\frac{a}{2}-
\frac{1}{4}}},
\end{equation}
\begin{equation}
\label{eq:9}
y_1=1+a\frac{x^2}{2!}+(a^2 +\frac{1}{2})\frac{x^4}{4!}+(a^3 +\frac{7a}{2})
\frac{x^6}{6!}+(a^4 +11a^2+\frac{15}{4})\frac{x^8}{8!} + \dots,
\end{equation}
\begin{equation}
\label{eq:10}
y_2=x+a\frac{x^3}{3!}+(a^2+\frac{3}{2})\frac{x^5}{5!}+(a^3+\frac{13a}{2})
\frac{x^7}{7!}+(a^4+17a^2+\frac{63}{4})\frac{x^9}{9!}+\dots,
\end{equation}
in which the coefficients $A_n$ of $\frac{x^n}{n!}$ obey the recurrence relation
\begin{equation}
\label{eq:11}
A_{n+2}=aA_n +\tfrac{1}{4}n(n-1)A_{n-2}.
\end{equation}
Similarly to Eq.~(\ref{eq:4}), there is relation
\begin{equation}
\label{eq:12}
U(a,x)=\frac{\pi}{\Gamma(a+\tfrac{1}{2})\cos^2 (\pi a)}[V(a,-x)-\sin(\pi a)V(a,x)].
\end{equation}
At $x=0$,
\begin{eqnarray} 
\label{eq:13}
U(a,0)=\frac{\sqrt\pi}{2^{\frac{a}{2}+\tfrac{1}{4}}\Gamma(\tfrac{a}{2}+\tfrac{3}{4})}, \nonumber \\
U^\prime(a,0)=-\frac{\sqrt\pi}{2^{\frac{a}{2}-\frac{1}{4}}\Gamma(\tfrac{a}{2}+\tfrac{1}{4})},
\end{eqnarray}
\begin{eqnarray}
\label{eq:14}
V(a,0)=\frac{2^{\frac{a}{2}+\frac{1}{4}}\sin\pi(\frac{3}{4}-\frac{a}{2})}
{\Gamma(\frac{3}{4}-\frac{a}{2})}, \nonumber \\
V^\prime(a,0)=\frac{2^{\frac{a}{2}+\frac{3}{4}}\sin\pi(\frac{1}{4}-\frac{a}{2})}
{\Gamma(\frac{1}{4}-\frac{a}{2})},
\end{eqnarray}

\subsection{Recurrence relations for $U(a,x)$ and $V(a,x)$}

Standard solutions $U(a,x)$ and $V(a,x)$ obey the recurrence relations
\begin{eqnarray}
\label{eq:15}
xU(a,x)-U(a-1,x)+(a+\tfrac{1}{2})U(a+1,x)=0, \nonumber \\
U^\prime(a,x)-\tfrac{1}{2}xU(a,x)+U(a-1,x)=0,
\end{eqnarray}
\begin{eqnarray}
\label{eq:16}
xV(a,x)-V(a+1,x)+(a-\tfrac{1}{2})V(a-1,x)=0, \nonumber \\
V^\prime(a,x)-\tfrac{1}{2}xV(a,x)-(a-\tfrac{1}{2})V(a-1,x)=0.
\end{eqnarray}

\subsection{Relations at large values of argument $x$}

At large values of argument $x$, when $x \gg |a|$, there are relations
\begin{equation}
\label{eq:17}
U(a,x)\sim x^{-a-\frac{1}{2}}e^{-\frac{x^2}{4}}\Big[1-\frac{(a+\frac{1}{2})
(a+\frac{3}{2})}{2x^2}+\frac{(a+\frac{1}{2})(a+\frac{3}{2})(a+\frac{5}{2})
(a+\frac{7}{2})}{2\cdot4x^4}-\dots\Big],
\end{equation}
\begin{equation}
\label{eq:18}
V(a,x) \sim \sqrt{\tfrac{2}{\pi}}
x^{a-\frac{1}{2}}e^{\frac{x^2}{4}}\Big[1+\frac{(a-\frac{1}{2})
(a-\frac{3}{2})}{2x^2}+\frac{(a-\frac{1}{2})(a-\frac{3}{2})(a-\frac{5}{2})
(a-\frac{7}{2})}{2\cdot4x^4}+\dots\Big].
\end{equation}

\subsection{Analytic relations at specific values of parameter $a$}

At half of odd parameter $a$, the standard solutions obey relations given in Table~\ref{tab:1}.
\begin{table}[h]
\begin{center}
\caption{\label{tab:1}Analytic relations for $U(a,x)$ and $V(a,x)$ at half of odd parameter $a$}
\begin{tabular}{|l|l|}
\hline
$U(-0.5,x)=e^{-\frac{x^2}{4}}$ & $V(0.5,x)=\sqrt{\tfrac{2}{\pi}}e^{\frac{x^2}{4}}$ \\
$U(-1.5,x)=x e^{-\frac{x^2}{4}}$ & $V(1.5,x)=\sqrt{\tfrac{2}{\pi}} x e^{\frac{x^2}{4}}$ \\
$U(-2.5,x)=(x^2-1)e^{-\frac{x^2}{4}}$ & $V(2.5,x)=\sqrt{\tfrac{2}{\pi}}(x^2+1)e^{\frac{x^2}{4}}$ \\
$U(-3.5,x)=(x^3-3x)e^{-\frac{x^2}{4}}$ & $V(3.5,x)=\sqrt{\tfrac{2}{\pi}}(x^3+3x)e^{\frac{x^2}{4}}$ \\
$U(-4.5,x)=(x^4-6x^2+3)e^{-\frac{x^2}{4}}$ & $V(4.5,x)=\sqrt{\tfrac{2}{\pi}}(x^4+6x^2+3)e^{\frac{x^2}{4}}$ \\
\hline
\end{tabular}
\end{center}
\end{table}

There is also relation,
\begin{equation}
\label{eq:19}
U(0.5,x)=\sqrt{\tfrac{\pi}{2}}e^{\frac{x^2}{4}} \mathrm{erfc} (\frac{x}{\sqrt 2}),
\end{equation}
where $\mathrm{erfc}(z)$ is the complementary error function~\cite{3}. Using recurrence relations~(\ref{eq:15}) and taking into account that
\begin{equation}
\label{eq:20}
\frac{\mathrm{d}~\mathrm{erfc}(\frac{x}{\sqrt 2})}
{\mathrm{d} x}=-\sqrt{\tfrac{2}{\pi}}e^{-\frac{x^2}{2}},
\end{equation}
one obtains
\begin{equation}
\label{eq:21}
U^\prime (0.5,x)=\frac{x}{2}U(0.5,x)-e^{-\frac{x^2}{4}}.
\end{equation}
Further we obtain
\begin{eqnarray}
\label{eq:22}
U(1.5,x)=-x U(0.5,x)+ e^{-\frac{x^2}{4}}, \nonumber \\
U(2.5,x)=\tfrac{1}{2}(x^2+1)U(0.5,x)-\tfrac{1}{2}x e^{-\frac{x^2}{4}},
\end{eqnarray} 
\begin{eqnarray}
\label{eq:23}
U^\prime (1.5,x)=-(\frac{x^2}{2}+1)U(0.5,x)+\frac{x}{2}e^{-\frac{x^2}{4}}, \nonumber \\
U^\prime (2.5,x)=\tfrac{1}{4}x(x^2+5)U(0.5,x)-(\frac{x^2}{4}+1)e^{-\frac{x^2}{4}}.
\end{eqnarray}
At integer values of parameter $a$, the standard solutions $U(a,x)$ and $V(a,x)$ obey relations given in Table~\ref{tab:2}.
\begin{table}[h]
\begin{center}
\caption{\label{tab:2}Analytic relations for $U(a,x)$ and $V(a,x)$ at integer values of parameter $a$}
\begin{tabular}{|l|l|}
\hline
$U(0,x)=\pi^{-\frac{1}{2}}(\frac{x}{2})^{\frac{1}{2}}K_{\frac{1}{4}}$ &
$V(0,x)=\tfrac{1}{2}\frac{x}{2}\mathfrak{I}_{\frac{1}{4}}$ \\
$U(1,x)=2\pi^{-\frac{1}{2}}(\frac{x}{2})^{\frac{3}{2}}(-K_{\frac{1}{4}}+K_{\frac{3}{4}})$ &
$V(1,x)=\tfrac{1}{2}\frac{x}{2}^{\frac{3}{2}}(\mathfrak{I}_{\frac{1}{4}}-\mathfrak{I}_{\frac{3}{4}})$ \\
$U(2,x)=\tfrac{4}{3}\pi^{-\frac{1}{2}}(\frac{x}{2})^{\frac{5}{2}}(2K_{\frac{1}{4}}-3K_{\frac{3}{4}}
+K_{\frac{5}{4}})$ &
$V(2,x)=\tfrac{1}{2}\frac{x}{2}^{\frac{5}{2}}(2\mathfrak{I}_{\frac{1}{4}}-3\mathfrak{I}_{\frac{3}{4}}+
\mathfrak{I}_{\frac{5}{4}})$ \\
\hline
$U(-1,x)=\pi^{-\frac{1}{2}}(\frac{x}{2})^{\frac{3}{2}}(K_{\frac{1}{4}}+K_{\frac{3}{4}})$ &
$V(-1,x)=\frac{x}{2}^{\frac{3}{2}}(\mathfrak{I}_{\frac{1}{4}}+\mathfrak{I}_{\frac{3}{4}})$ \\
$U(-2,x)=\pi^{-\frac{1}{2}}(\frac{x}{2})^{\frac{5}{2}}(2K_{\frac{1}{4}}+3K_{\frac{3}{4}}
-K_{\frac{5}{4}})$ &
$V(-2,x)=\tfrac{2}{3}\frac{x}{2}^{\frac{5}{2}}(2\mathfrak{I}_{\frac{1}{4}}+3\mathfrak{I}_{\frac{3}{4}}-
\mathfrak{I}_{\frac{5}{4}})$ \\
\hline
\end{tabular}
\end{center}
\end{table}
They are expressed in terms of modified Bessel functions $I_\nu(z)$ and $K_\nu(z)$, with
\begin{eqnarray}
\label{eq:24}
I_{-\nu}-I_\nu=\tfrac{2}{\pi}\sin(\pi\nu)K_\nu, \nonumber \\
I_{-\nu}+I_\nu=\cos(\pi\nu)\mathfrak{I}_\nu. 
\end{eqnarray}
The argument of all Bessel functions in Table~\ref{tab:2} is $\frac{x^2}{4}$. Further relations can be obtained by using the recurrence relations for the modified Bessel functions,
\begin{eqnarray}
\label{eq:25}
K_{\nu+1}(z)=K_{\nu-1}(z)+\frac{2\nu}{z} K_\nu (z) \nonumber \\
zK_\nu^\prime (z)=-z K_{\nu-1}(z)-\nu K_\nu(z),
\end{eqnarray}
\begin{eqnarray}
\label{eq:26}
I_{\nu+1} (z)=I_{\nu-1} (z) -\frac{2\nu}{z} I_\nu (z), \nonumber \\
zI_\nu^\prime (z)=z I_{\nu-1} (z)-\nu I_\nu (z),
\end{eqnarray}
with $K_{-\nu}(z)=K_\nu (z)$ and $\mathfrak{I}_{-\nu}(z)=\mathfrak{I}_\nu (z)$.

\section{Solutions of equation~\ref{eq:2}}

\subsection{Standard solution}

The standard solution $W(a,x)$ of Eq.~(\ref{eq:2}) is
\begin{equation}
\label{eq:27}
W(a,\pm x)=2^{-\frac{3}{4}}(\sqrt{\frac{G_1}{G_3}}y_1 \mp \sqrt{\frac{2G_3}{G_1}}y_2),
\end{equation}
where
\begin{equation}
\label{eq:28}
G_1=|\Gamma(\tfrac{ia}{2}+\tfrac{1}{4})|, \qquad G_3=|\Gamma(\tfrac{ia}{2}+\tfrac{3}{4})|,
\end{equation}
\begin{equation}
\label{eq:29}
y_1=1+a\frac{x^2}{2!}+(a^2-\frac{1}{2})\frac{x^4}{4!}+(a^3-\frac{7a}{2})
\frac{x^6}{6!}+(a^4-11a^2+\frac{15}{4})\frac{x^8}{8!}+\dots,
\end{equation}
\begin{equation}
\label{eq:30}
y_2=x+a\frac{x^3}{3!}+(a^2-\frac{3}{2})\frac{x^5}{5!}+(a^3-\frac{13a}{2})
\frac{x^7}{7!}+(a^4-17a^2+\frac{63}{4})\frac{x^9}{9!}+\dots,
\end{equation}
in which the coefficients $A_n$ of $\frac{x^n}{n!}$ obey the recurrence relation
\begin{equation}
\label{eq:31}
A_{n+2}=aA_n-\tfrac{1}{4}n(n-1)A_{n-2}.
\end{equation}
Relations for Gamma function of complex argument are given in Appendix~\ref{sec:app}.
At $x=0$,
\begin{eqnarray}
\label{eq:32}
W(a,0)=2^{-\frac{3}{4}}\sqrt{\frac{G_1}{G_3}}, \nonumber \\
W^\prime(a,0)=-2^{-\frac{1}{4}}\sqrt{\frac{G_3}{G_1}}.
\end{eqnarray}

\subsection{Relations at large values of argument $x$}

At large values of the argument $x$, when $x \gg |a|$, there are relations
\begin{eqnarray}
\label{eq:33}
W(a,x)=\sqrt\frac{2k}{x} [s_1(a,x)\cos\gamma-s_2(a,x)\sin\gamma], \nonumber \\
W(a,-x)=\sqrt\frac{2}{kx} [s_1(a,x)\sin\gamma+s_2(a,x)\cos\gamma],
\end{eqnarray}
where
\begin{equation}
\label{eq:34}
k=\sqrt{1+e^{2\pi a}}-e^{\pi a}, \qquad k^{-1}=\sqrt{1+e^{2\pi a}}+e^{\pi a},
\end{equation}
\begin{equation}
\label{eq:35}
\gamma=\frac{x^2}{4}-a\ln{x}+\frac{\pi}{4}+\frac{\phi}{2},
\end{equation}
with
\begin{equation}
\label{eq:36}
\phi=\arg{\Gamma(ia+\tfrac{1}{2})},
\end{equation}
\begin{equation}
\label{eq:37}
s_1(a,x)\sim 1+\frac{v_2}{1!2x^2}-\frac{u_4}{2!2^2x^4}-\frac{v_6}{3!2^3x^6}+
\frac{u_8}{4!2^4x^8}+\dots,
\end{equation}
\begin{equation}
\label{eq:38}
s_2(a,x) \sim -\frac{u_2}{1!2x^2}-\frac{v_4}{2!2^2x^4}+\frac{u_6}{3!2^3x^6}+
\frac{v_8}{4!2^4x^8}-\dots,
\end{equation}
with
\begin{equation}
\label{eq:39}
u_m+iv_m=\frac{\Gamma(m+ia+\frac{1}{2})}{\Gamma(ia+\frac{1}{2})},
\qquad m=2,4,6\dots
\end{equation}

\subsection{Analytic relations at $a=0$}

At $a=0$ there are relations
\begin{equation}
\label{eq:40}
W(0,\pm x)=2^{-\frac{5}{4}}\sqrt{\pi x}[J_{-\frac{1}{4}}(\frac{x^2}{4})\mp
J_{\frac{1}{4}}(\frac{x^2}{4})],
\end{equation}
\begin{equation}
\label{eq:41}
\frac{\mathrm{d}W(0,\pm x)}{\mathrm{d}x}=-2^{-\frac{9}{4}}x\sqrt{\pi x} [J_{\frac{3}{4}}(\frac{x^2}{4})\pm
J_{-\frac{3}{4}}(\frac{x^2}{4})],
\end{equation}
where $J_\nu(z)$ is the Bessel function of the first kind.

\section{Implementation of parabolic cylinder functions in Matlab}

Routines implemented in Matlab~\cite{4} for computation of parabolic cylinder functions  are shortly described in Table~\ref{tab:3}. For moderate values of argument $x$ and parameter $a$, standard parabolic cylinder functions $U(a,x)$ and $V(a,x)$ are computed with routines ``pu'' and ``pv'', respectively, whereas $W(a,x)$ is computed with routine ``pw''. Differentiation with respect to argument $x$ is computed with routines ``dpu'', ``dpv'', and ``dpw''. For large values of argument $x$, when $|x|\gg|a|$, functions 
$U(a,x)$, $V(a,x)$, and $W(a,x)$ are computed with routines ``pulx'', ``pvlx'', and ``pwlx'', and their derivatives with routines ``dpulx'', ``dpvlx'', and ``dpwlx'', respectively. Routine ``cgamma'' computes the gamma function of complex argument; using the function code $kf$, it computes either the logarithm of gamma function (when $kf=0$) or gamma function (when $kf=1$). Values of parabolic cylinder functions obtained by using these routines are shown in Tables~\ref{tab:4}--\ref{tab:9}.

\appendix
\section{\label{sec:app}Relations for gamma function of complex argument}

If $|z|\gg 1$ and $|\arg z|\leq \pi-\epsilon$ with $\epsilon >0$, there is relation
\begin{equation}
\label{eq:42}
\ln\Gamma(z)\sim(z-\tfrac{1}{2})\ln z -z+\tfrac{1}{2}\ln{2\pi}+\sum^\infty_{n=1}
\frac{B_{2n}}{2n(2n-1)}\frac{1}{z^{2n-1}},
\end{equation}
where $B_{2n}$ are the Bernoulli's numbers,
\begin{equation}
\label{eq:43}
B_{2k}=(-1)^{k-1}\frac{2\cdot(2k)!}{(2\pi)^{2k}}\sum^\infty_{n=1}\frac{1}{n^{2k}},
\qquad k=1,2,\dots
\end{equation}
Specific values, 
\begin{eqnarray}
\label{eq:44}
B_2&=1/6 \qquad &B_{12}=-691/2730 \nonumber \\
B_4&=-1/30 \qquad &B_{14}=7/6 \nonumber \\
B_6&=1/42 \qquad &B_{16}=-3617/510 \\
B_8&=-1/30 \qquad &B_{18}=43867/798 \nonumber \\
B_{10}&=5/66 \qquad &B_{20}=-174611/330 \nonumber 
\end{eqnarray}
Other useful relations,
\begin{eqnarray}
\label{eq:45}
\Gamma(z+n)=z(z+1)\cdots(z+n-1)\Gamma(z), \nonumber \\
\Gamma(z)\Gamma(-z)=\frac{-\pi}{z\sin(\pi z)}.
\end{eqnarray}

\begin{table}[ht]
\begin{center}
\caption{\label{tab:3}Routines for parabolic cylinder functions}
\begin{tabular}{|l|l|l|}
\hline
Name of routine & Routine call & What the routine computes \\
\hline          
cgamma&$[gr,gi]$=cgamma$(x,y,kf)$ &$\Gamma(z)$ with complex argument $z$ (when $kf=1$) or $\ln\Gamma(z)$, \\&& (when $kf=0$); $x$ and $y$ are the real and imaginary parts of $z$ \\&& $gr$ and $gi$ are the real and imaginary parts of $\ln\Gamma(z)$ or $\Gamma(z)$ \\ &&[Eqs.~(\ref{eq:42}--\ref{eq:45})]. \\
pu&$u$=pu$(a,x)$&Parabolic cylinder function $U(a,x)$ for moderate values of  \\
&&parameter $a$ and argument $x$ [Eqs.~(\ref{eq:3}--\ref{eq:12})]. \\
dpu&$du$=dpu$(a,x)$& Derivative with respect to $x$ of parabolic cylinder function   \\
&&$U(a,x)$ for moderate values of parameter $a$ and argument $x$ \\
pv&$v$=pv$(a,x)$&Parabolic cylinder function $V(a,x)$ for moderate values of   \\
&&parameter $a$ and argument $x$ [Eqs.~(\ref{eq:3}--\ref{eq:12})]. \\
dpv&$dv$=dpv$(a,x)$& Derivative with respect to $x$ of parabolic cylinder function   \\
&&$V(a,x)$ for moderate values of parameter $a$ and argument $x$ \\
pw&$w$=pw$(a,x)$&Parabolic cylinder function $W(a,x)$ for moderate values of   \\
&&parameter $a$ and argument $x$ [Eqs.~(\ref{eq:27}--\ref{eq:31})]. \\
dpw&$dw$=dpw$(a,x)$& Derivative with respect to $x$ of parabolic cylinder function  \\
&&$W(a,x)$ for moderate values of parameter $a$ and argument $x$ \\
pulx&$u$=pulx$(a,x)$&Parabolic cylinder function $U(a,x)$ for large values of parameter  \\
&&$x$ ($|x|\gg|a|$) and moderate values of parameter $a$ [Eq.~(\ref{eq:17})].\\
dpulx&$du$=dpulx$(a,x)$& Derivative with respect to $x$ of parabolic cylinder function  \\ 
&&$U(a,x)$ for large values of parameter $x$ ($|x|\gg|a|$) and moderate \\
&&values of parameter $a$.\\
pvlx&$v$=pvlx$(a,x)$&Parabolic cylinder function $V(a,x)$ for large values of parameter  \\
&&$x$ ($|x|\gg|a|$) and moderate values of parameter $a$ [Eq.~(\ref{eq:18})].\\
dpvlx&$dv$=dpvlx$(a,x)$& Derivative with respect to $x$ of parabolic cylinder function  \\ 
&&$V(a,x)$ for large values of parameter $x$ ($|x|\gg|a|$) and moderate \\
&&values of parameter $a$.\\
pwlx&$w$=pwlx$(a,x)$&Parabolic cylinder function $W(a,x)$ for large values of parameter \\
&&$x$ ($|x|\gg|a|$) and moderate values of parameter $a$ [Eqs.~(\ref{eq:33}--\ref{eq:39})].\\
dpwlx&$dw$=dpwlx$(a,x)$& Derivative with respect to $x$ of parabolic cylinder function  \\ 
&&$W(a,x)$ for large values of parameter $x$ ($|x|\gg|a|$) and moderate \\
&&values of parameter $a$.\\
\hline
\end{tabular}
\end{center}
\end{table}
\begin{table}[ht]
\begin{center}
\caption{\label{tab:4}Values of $U(a,x)$ with $a=-5,-3.5,-1,1,3.5,5$ and $x=0,1,3,5$} 
\begin{tabular}{|c|c|c|c|}
\hline
$x\backslash a$ & -5.0 & -3.5 & -1.0 \\
\hline          
 0.0 &  3.052183664350372 & -0.000000000000000 &  0.581368317019118 \\
 1.0 &  0.579926011661105 & -1.557601566142810 &  0.842203244069839 \\
 3.0 &  3.202129097812791 &  1.897186042113549 &  0.184881790005045 \\
 5.0 &  1.879976816310843 &  0.212349954984646 &  0.004337473181400 \\
\hline
$x\backslash a$ & 1.0 & 3.5 & 5.0\\
\hline          
 0.0 & 1.162736634038237 &  0.333333333333333 &  0.103354367470066 \\
 1.0 & 0.378262434740955 &  0.048971230815929 &  0.010659966828235 \\
 3.0 & 0.017224293634316 &  0.000610423938072 &  0.000070950238455 \\
 5.0 & 0.000161381143270 &  0.000002208878109 &  0.000000155227075 \\
\hline
\end{tabular}
\end{center}
\end{table}
\begin{table}[h]
\begin{center}
\caption{\label{tab:5}Values of $U(a,-x)$ with $a=-5,-3.5,-1,1,3.5,5$ and $x=0,1,3,5$} 
\begin{tabular}{|c|c|c|c|}
\hline
$x\backslash a$ & -5.0 & -3.5 & -1.0\\
\hline          
 0.0 &   3.052183664350372 & -0.000000000000000 &  0.581368317019118 \\
 1.0 &  -4.332232266251285 &  1.557601566142810 & -0.195001018223362 \\
 3.0 &   3.802753160685226 & -1.897186042113549 & -1.767855400724101 \\
 5.0 &  -9.615606269532364 & -0.212349954984649 &-35.754085404247576 \\
\hline
$x\backslash a$ & 1.0 & 3.5 & 5.0\\
\hline          
 0.0 &    1.16273663404 &      0.33333333333 &      0.10335436747 \\
 1.0 &    3.27078479478 &      2.19468750736 &      0.97838806074 \\
 3.0 &   45.73101176423 &    142.69397188181 &    125.30190015651 \\
 5.0 & 3259.12460949910 &  30297.53050402874 &  45998.28922772748 \\
\hline
\end{tabular}
\end{center}
\end{table}
\begin{table}[ht]
\begin{center}
\caption{\label{tab:6}Values of $V(a,x)$ with $a=-5,-3.5,-1,1,3.5,5$ and $x=0,1,3,5$}
\begin{tabular}{|c|c|c|c|}
\hline
$x\backslash a$ & -5.0 & -3.5 & -1.0\\
\hline          
 0.0 & -0.058311457540778 &  0.265961520267622 & -0.656003897333753 \\
 1.0 &  0.082766571619165 & -0.076762147625440 &  0.220035086525655 \\
 3.0 & -0.072650962016911 &  0.097154672861824 &  1.994811204614366 \\
 5.0 &  0.183704546768818 &  1.173350875864019 & 40.344165108706711 \\
\hline
$x\backslash a$ & 1.0 & 3.5 & 5.0 \\
\hline          
 0.0 &    0.3280019487 &      0             &      1.7220102305 \\
 1.0 &    0.9226713556 &      4.0980162226  &     16.3011422859 \\
 3.0 &   12.9004802412 &    272.5242458690  &   2087.6829809173 \\
 5.0 &  919.3820780818 &  57864.0209141053  & 766387.7838412275 \\
\hline
\end{tabular}
\end{center}
\end{table}
\begin{table}[h]
\begin{center}
\caption{\label{tab:7}Values of $V(a,-x)$ with $a=-5,-3.5,-1,1,3.5,5$ and $x=0,1,3,5$} 
\begin{tabular}{|c|c|c|c|}
\hline
$x\backslash a$ & -5.0 & -3.5 & -1.0\\
\hline          
 0.0 & -0.058311457540778 &  0.265961520267622 & -0.656003897333753 \\
 1.0 & -0.011079389291262 & -0.076762147625440 & -0.950324595068664 \\
 3.0 & -0.061176139925034 &  0.097154672861824 & -0.208616760217021 \\
 5.0 & -0.035916642101972 &  1.173350875864019 & -0.004894314375732 \\
\hline
$x\backslash a$ & 1.0 & 3.5 & 5.0 \\
\hline          
 0.0 &  0.32800194867 &       0             &  1.72201023050 \\
 1.0 &  0.10670586276 &      -4.09801622261 &  0.17760809131 \\
 3.0 &  0.00485888353 &    -272.52424586904 &  0.00118211779 \\
 5.0 &  0.00004552478 &  -57864.02091410524 &  0.00000258678 \\
\hline
\end{tabular}
\end{center}
\end{table}
\begin{table}[ht]
\begin{center}
\caption{\label{tab:8}Values of $W(a,x)$ with $a=-5,-3,-1,1,3,5$ and $x=0,1,3,5$} 
\begin{tabular}{|c|c|c|c|}
\hline
$x\backslash a$ & -5.0 & -3.0 & -1.0 \\
\hline          
 0.0 &  0.473478576486605 &  0.539330386270653 &  0.731481090245431 \\
 1.0 & -0.657520526362908 & -0.611126375982879 & -0.184115556183355 \\
 3.0 & -0.062604004232077 &  0.636305300554784 & -0.053352644054153 \\
 5.0 &  0.089361847055232 &  0.437066960213013 & -0.570254174032845 \\
\hline
$x\backslash a$ & 1.0 & 3.0 & 5.0\\
\hline          
 0.0 &  0.731481090245431 &  0.539330386270653 &  0.473478576486605 \\
 1.0 &  0.315937643962764 &  0.101682226485666 &  0.052572013487910 \\
 3.0 &  0.016773032899024 &  0.009166528652640 &  0.001223742332881 \\
 5.0 &  0.022807516888135 & -0.003844865237560 &  0.000115773464320 \\
\hline
\end{tabular}
\end{center}
\end{table}
\begin{table}[h]
\begin{center}
\caption{\label{tab:9}Values of $W(a,-x)$ with $a=-5,-3,-1,1,3,5$ and $x=0,1,3,5$} 
\begin{tabular}{|c|c|c|c|}
\hline
$x\backslash a$ & -5.0 & -3.0 & -1.0 \\
\hline          
 0.0 &  0.473478576486605 &  0.539330386270653 &  0.731481090245431 \\
 1.0 &  0.070610950611453 &  0.428801301530536 &  0.950916920458344 \\
 3.0 &  0.606270877302830 &  0.177268761402591 & -0.757374330077355 \\
 5.0 &  0.538608396875686 & -0.370945283780393 &  0.180907184885679 \\
\hline
$x\backslash a$ & 1.0 & 3.0 & 5.0 \\
\hline          
 0.0 &  0.731481090245 &    0.539330386271 &    0.473478576487 \\
 1.0 &  1.903689596383 &    3.001251077335 &    4.378212848013 \\
 3.0 &  6.183176599808 &   57.210355295947 &  253.398744868662 \\
 5.0 & -4.359927574948 &   66.590129609337 & 2852.835947866653 \\
\hline
\end{tabular}
\end{center}
\end{table}
 
\end{document}